\documentclass[seceq,preprint]{ptptex}
\usepackage{wrapft}
\usepackage{graphicx}

\newcommand{\bra}[1]{\langle {#1} |}     
\newcommand{\ket}[1]{| {#1} \rangle}     
\newcommand{\bbra}[1]{\langle\!\langle {#1} |}     
\newcommand{\kket}[1]{| {#1} \rangle\!\rangle}     
\newcommand{\rbra}[1]{( {#1} |}     
\newcommand{\rket}[1]{| {#1} )}     
\newcommand{\wtilde}[1]{\widetilde{#1}} 
\newcommand{\ovl}[1]{\overline{#1}} 

\def\<{\langle}
\def\>{\rangle}
\def\bsub{\begin{subequations}}
\def\esub{\end{subequations}}
\def\beqn{\begin{eqnarray}}
\def\eeqn{\end{eqnarray}}
\def\b{\begin{equation}}

\title{
Boson Realization of the $su(3)$-Algebra. II
}
\subtitle{
Holstein-Primakoff Representation for the Lipkin Model
}    

\author{
Constan\c{c}a {\sc Provid\^encia},$^{1}$
Jo\~ao da {\sc Provid\^encia},$^{1}$\\
Yasuhiko {\sc Tsue}$^{2}$ 
and Masatoshi {\sc Yamamura}$^{3}$
}

\inst{
$^{1}$Departamento de Fisica, Universidade de Coimbra, 3004-516 Coimbra, 
Portugal\\
$^2$Physics Division, Faculty of Science, Kochi University, Kochi 780-8520, 
Japan\\
$^{3}$Faculty of Engineering, Kansai University, Suita 564-8680, Japan
}

\recdate{
\today
}

\abst{
On the basis of the Schwinger boson representation for the Lipkin model 
developed in (I), 
the Holstein-Primakoff representation for the 
$su(3)$-algebra is presented. 
Including the symmetric case, the representation obtained in this paper 
contains all the cases. 
}
\begin{document}
\maketitle

\section{Introduction}

It is well known that we have tow forms for the boson realization of the 
$su(2)$-algebra. 
One is called the Schwinger boson representation\cite{1} and the 
other the Holstein-Primakoff boson representation.\cite{2} 
Each has its own characteristic and merit. 
In the Schwinger representation, the $su(2)$-generators can be expressed 
in terms of bilinear form with respect to two kinds of boson operators. 
In the Holstein-Primakoff representation, they are expressed in terms of 
one kind of boson operator with the square root type operator.

When we describe fluctuation around the equilibrium, boson operators are 
helpful. 
In the case of the Holstein-Primakoff representation, the equailibrium 
is introduced from the out-side and the boson operator plays a role of 
describing only fluctuation. 
Contrarily, in the case of the Schwinger representation, the equilibrium 
itself is treated by boson operators, and then, the separation of 
the boson operators into two roles may be devised in each problem 
under investigation not only for the equilibrium but also for the 
fluctuation. 
Therefore, if the flcutuation is not so large, the Holstein-Primakoff 
representation permits us to describe the fluctuation more transparently than 
the case of the Schwinger boson representation.

In the previous paper,\cite{3} which, hereafter, is referred to as (I), 
we presented the Schwinger boson representation for the 
$su(3)$-algebra.\cite{3} 
This form is a concrete development of the $su(3)$-algebra in the case of 
the $su(M+1)$-algebra for the Lipkin model.\cite{4} 
With the aid of the framework given in this form, we can describe the 
$su(3)$-Lipkin model in terms of the Schwinger boson representation. 
From the reason mentioned in the introductory part of this section, 
it may be interesting to present the Holstein-Primakoff representation in 
a complete form. 
The case of the symmetric representation is well known.\cite{5}

A main aim of this paper is to present the Holstein-Primakoff representation 
of the $su(3)$-algebra as a disguised form of the Schwinger representation 
developed in (I). 
Usually, the Hosltein-Primakoff representation is obtained in terms of 
certain ordering of boson operators after applying the MYT boson mapping 
method proposed by Marumori, Yamamura and Tokunaga.\cite{6} 
This idea was demonstrated by Marshalek.\cite{5} 
However, in this paper, we adopt another idea. 
Regarding the intrinsic state in the Schwinger representation as 
representing the equilibrium, the fluctuations around the equilibrium 
are described by boson operators. 
In the case of the $su(2)$-algebra, the intrinsic state is specified 
by one quantum number which expresses the magnitude of the 
$su(2)$-spin. 
Then, the Holstein-Primakoff representation of the $su(2)$-algebra is 
characterized by the magnitude of the $su(2)$-spin. 
In the case of the $su(3)$-algebra, the intrinsic state is characterized 
by two quantum numbers, and then, the Holstein-Primakoff representation is 
characterized by two quantum numbers. 
Of course, the $su(3)$-generators are expressed in terms of these two 
quantum numbers.

In \S 2, the basic idea is demonstrated in the case of well-known 
$su(2)$-algebra. 
Section 3 is a central part of this paper. 
Following basic idea developed in \S 2, the Holstein-Primakoff representation 
for the $su(3)$-algebra is presented. 
In \S 4, various properties of the representation obtained in \S 3 are 
discussed. 
Especially, the physical space, which is a certain subspace in the 
whole boson space, is given. 
Finally, the case of the simplest approximation is discussed in 
relation to the RPA method in the fermion space.

\section{An illustrative example for the basic idea --- the case of the 
$su(2)$-algebra ---}

For the preparation for our final aim, we will sketch the case of the 
$su(2)$-algebra. 
Through this sketch, we can demonstrate our basic viewpoint. 
The $su(2)$-algebra consisits of three generators ${\hat S}_{\pm,0}$, 
which can be expressed in terms of two kinds of bosons 
$({\hat a} , {\hat a}^*)$ and $({\hat b} , {\hat b}^*)$: 
\begin{equation}\label{2-1}
{\hat S}_+={\hat a}^*{\hat b} \ , \qquad 
{\hat S}_-={\hat b}^*{\hat a} \ , \qquad
{\hat S}_0=(1/2)({\hat a}^*{\hat a}-{\hat b}^*{\hat b}) \ . 
\end{equation}
In the representation (\ref{2-1}), an operator which commutes with 
${\hat S}_{\pm,0}$ can be defined as follows:
\begin{equation}\label{2-2}
{\hat S}=(1/2)({\hat a}^*{\hat a}+{\hat b}^*{\hat b}) \ . 
\end{equation}
The intrinsic state $\ket{s}$ obeys the condition 
\begin{equation}\label{2-3}
{\hat S}_-\ket{s}=0 \ , \qquad 
{\hat S}_0\ket{s}=-s\ket{s} \ . 
\end{equation}
Explicitly, $\ket{s}$ is given as 
\begin{equation}\label{2-4}
\ket{s}=\left(\sqrt{(2s)!}\right)^{-1}({\hat b}^*)^{2s}\ket{0} \ . 
\qquad (s=0,\ 1/2,\ 1,\cdots )
\end{equation}
In (A), we used the notation 
\begin{equation}\label{2-5}
m_0=2s\ . \qquad (m_0=0,\ 1,\ 2, \cdots)
\end{equation}
Since ${\hat S}_+$ plays a role of the excited state generating operator, the 
excited state denoted as $\ket{s,s_0}$ is obtained in the form 
\begin{equation}\label{2-6}
\ket{s,s_0}=\sqrt{\frac{1}{(2s)!}\frac{(s-s_0)!}{(s+s_0)!}}
({\hat S}_+)^{s+s_0}\ket{s} \ . \qquad
(s_0=-s,\ -s+1,\ \cdots,\ s-1,\ s)
\end{equation}
Operation of ${\hat S}$ on the state $\ket{s,s_0}$ gives us 
\begin{equation}\label{2-7}
{\hat S}\ket{s,s_0}=s\ket{s,s_0} \ . 
\end{equation}
The definition of ${\hat S}$ leads us 
\bsub\label{2-8}
\beqn
& &({\hat a}^*{\hat a}+{\hat b}^*{\hat b})\ket{s,s_0}
=2s\ket{s,s_0} \ , 
\label{2-8a}
\eeqn
i.e., 
\beqn
& &{\hat b}^*{\hat b}\ket{s,s_0}=\left(
\sqrt{2s-{\hat a}^*{\hat a}}\right)^2 \ket{s,s_0} \ . 
\label{2-8b}
\eeqn
\esub
From the relation (\ref{2-8}), we have 
\begin{equation}\label{2-9}
{\hat b}^*{\hat b}\ket{c(m_0)}=\left(\sqrt{m_0-{\hat a}^*{\hat a}}\right)^2
\ket{c(m_0)}\ . 
\end{equation}
Here, $\ket{c(m_0)}$ is an arbitrary superposition of the states 
$\ket{s,s_0}$ $(s_0=-s,-s+1,\cdots ,s-1,s)$:
\begin{equation}\label{2-10}
\ket{c(m_0)}=\sum_{s_0}c_{s_0}(s)\ket{s,s_0} \ . \qquad
(m_0=2s) 
\end{equation}

For the relation (\ref{2-9}), we can give the following interpretation: 
The intrinsic state $\ket{s}$, which is expressed only in terms of the 
boson ${\hat b}^*$, plays the same role as that in the free vacuum 
of the Hartree-Fock theory $\ket{f}$. 
It shows the equilibrium which is characterized by the eigenvalue of 
the density matrix, i.e., $\rho_n=1$ and 0. 
The state $\ket{s}$ is characterized by the eigenvalue of ${\hat b}^*{\hat b}$, 
i.e., $2s$. 
Fluctuations around $\ket{f}$ can be described in terms of the operations 
of the particle-hole pairs on $\ket{f}$ and for the sake of the 
fluctuations, the value of $\rho_n$ decreases from 1. 
In the present system, fluctuations around $\ket{s}$ can be described 
in the framework of the operations of ${\hat S}_+$ on $\ket{s}$, i.e., 
the operations of ${\hat a}^*$ and ${\hat b}$. 
The operations of ${\hat a}^*$ make the fluctuations increase and 
the operations of ${\hat b}$ make the effect of $m_0$ decrease. 
The relation (\ref{2-9}) teaches us the above interpretation.

Concerning the above interpretation, the Holstein-Primakoff representation 
may be superior to the Schwinger representation. 
We introduce a new boson space in which, in this paper, the boson operator 
is denoted as $({\hat \alpha}, {\hat \alpha}^*)$ and the vacuum $\rket{0}$ 
plays the same role as that of $\ket{s}$ in the Schwinger representation. 
The fluctuations around $\rket{0}$ can be described in terms of the 
operations of ${\hat \alpha}^*$ and the decrease from the value $m_0$ 
is given by the operations of the operator 
$\sqrt{m_0-{\hat \alpha}^*{\hat \alpha}}$. 
This is suggested by the relation (\ref{2-9}). 
Then, the correspondence to the Schwinger representation is 
summarized as follows: 
\bsub\label{2-11}
\beqn
& &\ket{s} \sim \rket{0} \ , 
\label{2-11a}\\
& &({\hat a},{\hat a}^*) \sim ({\hat \alpha}, {\hat \alpha}^*) \ , 
\label{2-11b}\\
& &{\hat b}\ {\hbox{\rm and}}\ {\hat b}^* \sim 
\sqrt{m_0-{\hat \alpha}^*{\hat \alpha}} \ . \qquad (m_0=2s)
\label{2-11c}
\eeqn
\esub
Under the correspondence (\ref{2-11}), the form (\ref{2-1}) is replaced with 
the form 
\beqn\label{2-12}
& &{\hat S}_+ \sim {\ovl S}_+
={\hat \alpha}^*\sqrt{m_0-{\hat \alpha}^*{\hat \alpha}} \ , \nonumber\\
& &{\hat S}_- \sim {\ovl S}_-
=\sqrt{m_0-{\hat \alpha}^*{\hat \alpha}}\ {\hat \alpha} \ , \nonumber\\
& &{\hat S}_0 \sim {\ovl S}_0
={\hat \alpha}^*{\hat \alpha}-m_0/2 \ . 
\eeqn
The above is identical with the Holstein-Primakoff representation 
and we can prove 
\begin{equation}\label{2-13}
[\ {\ovl S}_+ \ , \ {\ovl S}_- \ ]=2{\ovl S}_0 \ , \qquad
[\ {\ovl S}_0 \ , \ {\ovl S}_\pm \ ]=\pm{\ovl S}_\pm \ . 
\end{equation}
In the strict sense, the form (\ref{2-12}) is valid in the subspace 
spanned by 
\begin{equation}\label{2-14}
\rket{n}=\left(\sqrt{n!}\right)^{-1}({\hat \alpha}^*)^n \rket{0} \ . \qquad
(n=0,\ 1,\ \cdots ,\ m_0)
\end{equation}
The space spanned by $\{\rket{n}; n=0,1,\cdots,m_0\}$ is called the 
physical space.

\section{The case of the $su(3)$-algebra}

Following the basic viewpoint for constructing the Holstein-Primakoff 
representation from the Schwinger representation, we present the form 
of the $su(3)$-algebra. 
First, we note the relations (I$\cdot$6$\cdot$4a) and 
(I$\cdot$6$\cdot$4b): 
\bsub\label{3-1}
\beqn
& &({\hat b}^*{\hat b}+{\hat a}_+^*{\hat a}_+
+{\hat a}_-^*{\hat a}_-) \ket{I^1I^0,II_0;T}=m_0\ket{I^1I^0,II_0;T} \ , 
\label{3-1a}\\
& &({\hat a}^*{\hat a}+{\hat b}_+^*{\hat b}_+
+{\hat b}_-^*{\hat b}_-) \ket{I^1I^0,II_0;T}=m_1\ket{I^1I^0,II_0;T} \ . 
\label{3-1b}
\eeqn
\esub
The relation (\ref{3-1}) corresponds to the relation (\ref{2-8a}) in 
the $su(2)$-algebra. 
Taking into account that the intrinsic state is expressed in terms
of ${\hat b}^*$ and ${\hat b}_-^*$, we rewrite the relation (\ref{3-1}) 
in the form 
\bsub\label{3-2}
\beqn
& &{\hat b}^*{\hat b} \ket{I^1I^0,II_0;T}=
\left(\sqrt{m_0-{\hat a}_+^*{\hat a}_+
-{\hat a}_-^*{\hat a}_-}\right)^2\ket{I^1I^0,II_0;T} \ , 
\label{3-2a}\\
& &{\hat b}_-^*{\hat b}_- \ket{I^1I^0,II_0;T}=
\left(\sqrt{m_1-{\hat a}^*{\hat a}
-{\hat b}_+^*{\hat b}_+}\right)^2\ket{I^1I^0,II_0;T} \ . 
\label{3-2b}
\eeqn
\esub
Of course, the relation (\ref{3-2}) gives us 
\bsub\label{3-3}
\beqn
& &{\hat b}^*{\hat b} \ket{c(m_0, m_1)}=
\left(\sqrt{m_0-{\hat a}_+^*{\hat a}_+
-{\hat a}_-^*{\hat a}_-}\right)^2\ket{c(m_0, m_1)} \ , 
\label{3-3a}\\
& &{\hat b}_-^*{\hat b}_- \ket{c(m_0, m_1)}=
\left(\sqrt{m_1-{\hat a}^*{\hat a}
-{\hat b}_+^*{\hat b}_+}\right)^2\ket{c(m_0, m_1)} \ . 
\label{3-3b}
\eeqn
\esub
Here, $\ket{c(m_0, m_1)}$ denotes an arbitrary superposition of the state 
$\ket{I^1I^0,II_0;T}$: 
\begin{equation}\label{3-4}
\ket{c(m_0, m_1)}=\sum_{I^1II_0}
C_{I^1II_0}(I^0,T)\ket{I^1I^0,II_0;T} \ . 
\end{equation}
It should be noted that, as was shown in the relation (I$\cdot$6$\cdot$2), 
$(m_0,m_1)$ is related to $(T,I^0)$ as follows: 
\begin{equation}\label{3-5}
m_0=2[(T-3/2)-I^0] \ , \qquad m_1=2I^0 \ . 
\end{equation}
The relation (\ref{3-3}) parallels the relation (\ref{2-9}) in the 
interpretation. 
The fluctuations around the state $\ket{m_0,m_1}$ can be described 
in terms of the boson operators $({\hat a}_+, {\hat a}_+^*)$, 
$({\hat a}_-, {\hat a}_-^*)$, $({\hat a}, {\hat a}^*)$ and 
$({\hat b}_+, {\hat b}_+^*)$, and by the fluctuations, the effects of 
$m_0$ 
and $m_1$ decrease.

Under the above argument, we complete the Holstein-Primakoff 
representation for the $su(3)$-algebra. 
For this purpose, a new boson space is introduced in terms of bosons 
$({\hat \alpha}_+, {\hat \alpha}_+^*)$, 
$({\hat \alpha}_-, {\hat \alpha}_-^*)$, $({\hat \alpha}, {\hat \alpha}^*)$ 
and $({\hat \beta}_+, {\hat \beta}_+^*)$. 
The vacuum $\rket{0}$ of this boson space plays the same role as that of 
$\ket{m_0,m_1}$. 
In the same way as that in \S 2, we set up the correspondence 
\bsub\label{3-6}
\beqn
& &\ket{m_0,m_1} \sim \rket{0} \ , 
\label{3-6a}\\
& &({\hat a}_+,{\hat a}_+^*) \ , \ ({\hat a}_-,{\hat a}_-^*)\ , \ 
({\hat a},{\hat a}^*) \ {\rm and} \ ({\hat b}_+,{\hat b}_+^*) \nonumber\\
& &\qquad 
 \sim 
 ({\hat \alpha}_+, {\hat \alpha}_+^*)\ , \ 
({\hat \alpha}_-, {\hat \alpha}_-^*)\ , \ 
({\hat \alpha}, {\hat \alpha}^*) \ {\rm and}\  
({\hat \beta}_+, {\hat \beta}_+^*) \ , 
\label{3-6b}\\
& &{\hat b}\ {\hbox{\rm and}}\ {\hat b}^* \sim 
\sqrt{m_0-{\hat \alpha}_+^*{\hat \alpha}_+-{\hat \alpha}_-^*{\hat \alpha}_-} 
\ ,  \nonumber\\
& &{\hat b}_-\ {\hbox{\rm and}}\ {\hat b}_-^* \sim 
\sqrt{m_1-{\hat \alpha}^*{\hat \alpha}-{\hat \beta}_+^*{\hat \beta}_+}  \ . 
\label{3-6c}
\eeqn
\esub
The expression (I$\cdot$2$\cdot$10) gives us the following form for the 
$su(3)$-generators: 
\bsub\label{3-7}
\beqn
& &{\hat I}_+ \sim {\ovl I}_+
={\hat \alpha}_+^*{\hat \alpha}_--{\hat \beta}_+^*
\sqrt{m_1-{\hat \alpha}^*{\hat \alpha}-{\hat \beta}_+^*{\hat \beta}_+} \ , 
\nonumber\\
& &{\hat I}_- \sim {\ovl I}_-
={\hat \alpha}_-^*{\hat \alpha}_+ -
\sqrt{m_1-{\hat \alpha}^*{\hat \alpha}-{\hat \beta}_+^*{\hat \beta}_+}\ 
{\hat \beta}_+ \ , 
\nonumber\\
& &{\hat I}_0 \sim {\ovl I}_0
={\hat \beta}_+^*{\hat \beta}_+
+(1/2)({\hat \alpha}^*{\hat \alpha}+{\hat \alpha}_+^*{\hat \alpha}_+
-{\hat \alpha}_-^*{\hat \alpha}_-)-m_1/2 \ , 
\label{3-7a}\\
& &{\hat M}_0 \sim {\ovl M}_0
=(3/2)({\hat \alpha}^*{\hat \alpha}+{\hat \alpha}_+^*{\hat \alpha}_+
+{\hat \alpha}_-^*{\hat \alpha}_-)-(m_0+m_1/2) \ , 
\label{3-7b}\\
& &{\hat D}_-^* \sim {\ovl D}_-^*
={\hat \alpha}_-^*
\sqrt{m_0-{\hat \alpha}_+^*{\hat \alpha}_+
-{\hat \alpha}_-^*{\hat \alpha}_-} +{\hat \alpha}^*{\hat \beta}_+ \ , 
\nonumber\\
& &{\hat D}_+^* \sim {\ovl D}_+^*
={\hat \alpha}_+^*
\sqrt{m_0-{\hat \alpha}_+^*{\hat \alpha}_+
-{\hat \alpha}_-^*{\hat \alpha}_-} 
+{\hat \alpha}^*
\sqrt{m_1-{\hat \alpha}^*{\hat \alpha}-{\hat \beta}_+^*{\hat \beta}_+} \ , 
\nonumber\\
& &{\hat D}_- \sim {\ovl D}_-
=\sqrt{m_0-{\hat \alpha}_+^*{\hat \alpha}_+
-{\hat \alpha}_-^*{\hat \alpha}_-}\ {\hat \alpha}_- 
+{\hat \beta}_+^*{\hat \alpha} \ , 
\nonumber\\
& &{\hat D}_+ \sim {\ovl D}_+
=\sqrt{m_0-{\hat \alpha}_+^*{\hat \alpha}_+
-{\hat \alpha}_-^*{\hat \alpha}_-} \ {\hat \alpha}_+
+\sqrt{m_1-{\hat \alpha}^*{\hat \alpha}-{\hat \beta}_+^*{\hat \beta}_+} \ 
{\hat \alpha}\ . 
\label{3-7c}
\eeqn
\esub
The $su(1,1)$-generators (I$\cdot$2$\cdot$11) and ${\wtilde R}_0$ shown 
in the relation (I$\cdot$2$\cdot$14) lead to 
\bsub\label{3-8}
\beqn
& &{\wtilde T}_+ \sim {\check T}_+
={\hat \alpha}^*
\sqrt{m_0-{\hat \alpha}_+^*{\hat \alpha}_+
-{\hat \alpha}_-^*{\hat \alpha}_-} 
-{\hat \alpha}_+^*
\sqrt{m_1-{\hat \alpha}^*{\hat \alpha}-{\hat \beta}_+^*{\hat \beta}_+}
-{\hat \alpha}_-^*{\hat \beta}_+^* \ , 
\nonumber\\
& &{\wtilde T}_- \sim {\check T}_-
=\sqrt{m_0-{\hat \alpha}_+^*{\hat \alpha}_+
-{\hat \alpha}_-^*{\hat \alpha}_-} \ {\hat \alpha}
-\sqrt{m_1-{\hat \alpha}^*{\hat \alpha}-{\hat \beta}_+^*{\hat \beta}_+}
\ {\hat \alpha}_+
-{\hat \beta}_+{\hat \alpha}_- \ , \qquad
\label{3-8a}\\
& &{\wtilde T}_0 \sim {\check T}_0
=(1/2)(m_1+m_0+3) \ , 
\label{3-8b}\\
& &{\wtilde R}_0 \sim {\check R}_0
=(1/2)(m_1-m_0) \ . 
\label{3-9}
\eeqn
\esub
Thus, we obtained the Holstein-Primakoff representation of 
the $su(3)$-algebra. 
In the next section, we will discuss various features of the representation.

\section{Various properties of the representation}

In the previous section, we developed the Holstein-Primakoff representation 
of the $su(3)$-algebra. 
As can be seen in the form (\ref{3-7}), it may be better to call the 
mixed representation. 
First of all, we must stress that the representation (\ref{3-7}) 
satisfies the commutation relations of the $su(3)$-algebra, which are listed 
in the relation (I$\cdot$2$\cdot$2). 
Through the straightforward calculation, we can confirm it. 
However, it should be noted that the expression (\ref{3-8}) for 
$({\wtilde T}_{\pm,0})$ does not satisfy the commutation relation 
for the $su(1,1)$-algebra. 
The forms (\ref{3-8b}) and (\ref{3-9}) are natural, because we construct 
the form (\ref{3-7}) under the condition that the eigenvalues of 
${\wtilde T}_0$ and ${\wtilde R}_0$ are $(1/2)(m_1+m_0+3)$ 
and $(1/2)(m_1-m_0)$, respectively. 
But, we will show the importance of the expression (\ref{3-8a}) for defining 
the physical space. 

Next, our interest is related to the physical space. 
First, we note the existence of the operators 
$\sqrt{m_0-{\hat \alpha}_+^*{\hat \alpha}_+-{\hat \alpha}_-^*{\hat \alpha}_-}$ 
and 
$\sqrt{m_1-{\hat \alpha}^*{\hat \alpha}-{\hat \beta}_+^*{\hat \beta}_+}$, 
which means that 
$(m_0-{\hat \alpha}_+^*{\hat \alpha}_+-{\hat \alpha}_-^*{\hat \alpha}_-)$ 
and $(m_1-{\hat \alpha}^*{\hat \alpha}-{\hat \beta}_+^*{\hat \beta}_+)$ 
should be positive definite. 
Then, the physical space should be spanned by the states with the 
condition 
\begin{equation}\label{4-1}
\hbox{\rm the\ boson\ number\ of\ }
\begin{cases} 
{\hat \alpha}_{+}^*{\hat \alpha}_+ + {\hat \alpha}_-^*{\hat \alpha}_-
=0, 1, 2, \cdots , m_0 \ , \\
{\hat \alpha}^*{\hat \alpha} + {\hat \beta}_+^*{\hat \beta}_+
=0, 1, 2, \cdots , m_1 \ .
\end{cases}
\end{equation}
From the reason mentioned below, only the condition (\ref{4-1}) is not 
sufficient for determining the physical space. 
The states, which compose the physical space, correspond to the states 
$\ket{I^1I^0,II_0;T}$ shown in the relation (I$\cdot$4$\cdot$4), i.e., 
\begin{equation}\label{4-2}
\ket{I^1I^0,II_0;T} \sim \rket{I^1I^0,II_0;T} \ . 
\end{equation}
Under the correspondences (\ref{3-6a}) and (\ref{3-7}), the state 
$\rket{I^1I^0,II_0;T}$ is obtained. 
In order to show $(m_0,m_1)$ explicitly, hereafter, $\rket{I^1I^0,II_0;T}$ 
is abbreviated as 
\begin{equation}\label{4-3}
\rket{I^1I^0,II_0;T} = \rket{I^1II_0;m_0m_1} \ . 
\end{equation}
Our present boson space consists of four kinds of the boson operators. 
However, even if obeying the condition (\ref{4-1}), 
$\rket{I^1II_0;m_0m_1}$ is expressed in terms of three quantum numbers, 
$(I^1,I,I_0)$. 
Therefore, in order to specify the physical space definitely, 
a condition additional to restriction (\ref{4-1}) is necessary. 
In order to find this condition, we pay attention to the form 
(\ref{3-8a}), from which the following relation is derived: 
\begin{equation}\label{4-4}
{\check T}_-\rket{I^1II_0;m_0m_1}=0 \ . 
\end{equation}
In the space where two operators 
$(\!\!\sqrt{\!m_0\!-\!{\hat \alpha}_+^*{\hat \alpha}_+\!-\!
{\hat \alpha}_-^*{\hat \alpha}_-}\!
)^{-1}$ and 
$(\!\!\sqrt{m_1\!-\!{\hat \alpha}^*{\hat \alpha}\!-\!
{\hat \beta}_+^*{\hat \beta}_+}\!
)^{-1}$ can be defined, we are able to prove the relations 
\bsub\label{4-5}
\beqn
& &[\ {\check \tau}_- \ , \ {\ovl I}_+ \ ]
=[\ {\check \tau}_- \ , \ {\ovl D}_\pm^* \ ]=0 \ , 
\label{4-5a}\\
& &{\check \tau}_-\ket{m_0,m_1}=0 \ . 
\label{4-5b}
\eeqn
\esub
Here, ${\check \tau}_-$ is defined as 
\begin{equation}\label{4-6}
{\check \tau}_-
=\left(
\sqrt{m_0-{\hat \alpha}_+^*{\hat \alpha}_+-{\hat \alpha}_-^*{\hat \alpha}_-}
\sqrt{m_1-{\hat \alpha}^*{\hat \alpha}-{\hat \beta}_+^*{\hat \beta}_+}
\right)^{-1}{\check T}_- \ . 
\end{equation}
With the use of the relation (\ref{4-5}), we have 
\beqn\label{4-7}
{\check T}_-\rket{I^1II_0;m_0m_1}
&=&\sqrt{m_0-{\hat \alpha}_+^*{\hat \alpha}_+-{\hat \alpha}_-^*{\hat \alpha}_-}
\sqrt{m_1-{\hat \alpha}^*{\hat \alpha}-{\hat \beta}_+^*{\hat \beta}_+}
\nonumber\\
& &\times{\check \tau}_-\rket{I^1II_0;m_0m_1}=0 \ . 
\eeqn
The relation (\ref{4-4}) gives us 
\bsub\label{4-8}
\begin{equation}\label{4-8a}
\rbra{I^1II_0;m_0m_1}{\check T}_+=0 \ , 
\end{equation}
namely, 
\begin{equation}\label{4-8b}
\rbra{I^1II_0;m_0m_1}({\check T}_+)^n=0 \ . \qquad (n=1,2,3,\cdots) 
\end{equation}
\esub
Then, we have 
\beqn
& &\rbra{I^1II_0;m_0m_1}nI^{1'}I'I_0';m_0m_1)=0 \ , 
\label{4-8-2}\\
& &\rket{nI^{1'}I'I_0';m_0m_1}=({\check T}_+)^n\rket{I^{1'}I'I_0';m_0m_1} \ .
\label{4-9}
\eeqn
Since ${\check T}_+$ is also defined in the space obeying the 
condition (\ref{4-1}), $n$ can run in the region governed by the condition. 
Further, we can prove 
\begin{equation}\label{4-10}
{\check T}_-\rket{nI^{1'}I'I_0';m_0m_1}\neq 0\ . \quad (n\neq 0)
\end{equation}
The above consideration gives us the conclusion that the physical space 
is defined as follows: 
In all the states composing the bosons ${\hat \alpha}^*$, ${\hat \alpha}_+^*$, 
${\hat \alpha}_-^*$ and ${\hat \beta}_+^*$, the state $\rket{ph}$ belonging 
to the physical space obeys the condition (\ref{4-1}), and further, the 
condition ${\check T}_-\rket{ph}=0$. 

Finally, we will discuss the symmetric representation. 
In the case $m_1=0$, 
$\sqrt{m_1-{\hat \alpha}^*{\hat \alpha}-{\hat \beta}_+^*{\hat \beta}_+}$ is 
defined only in the condition ${\hat \alpha}^*{\hat \alpha}=
{\hat \beta}_+^*{\hat \beta}_+=0$. 
Then, the form (\ref{3-7}) reduces to the following: 
\bsub\label{4-11}
\beqn
& &{\ovl I}_+={\hat \alpha}_+^*{\hat \alpha}_- \ , \qquad
{\ovl I}_-={\hat \alpha}_-^*{\hat \alpha}_+ \ , \qquad
{\ovl I}_0=(1/2)({\hat \alpha}_+^*{\hat \alpha}_+
-{\hat \alpha}_-^*{\hat \alpha}_-) \ , 
\label{4-11a}\\
& &{\ovl M}_0=(3/2)({\hat \alpha}_+^*{\hat \alpha}_+
+{\hat \alpha}_-^*{\hat \alpha}_-)-m_0 \ . 
\label{4-11b}\\
& &{\ovl D}_-^*={\hat \alpha}_-^*\sqrt{m_0-{\hat \alpha}_+^*{\hat \alpha}_+
-{\hat \alpha}_-^*{\hat \alpha}_-}\ , \quad
{\ovl D}_+^*={\hat \alpha}_+^*\sqrt{m_0-{\hat \alpha}_+^*{\hat \alpha}_+
-{\hat \alpha}_-^*{\hat \alpha}_-}\ , \nonumber\\
& &{\ovl D}_-=\sqrt{m_0-{\hat \alpha}_+^*{\hat \alpha}_+
-{\hat \alpha}_-^*{\hat \alpha}_-}\ {\hat \alpha}_-\ , \quad
{\ovl D}_+=\sqrt{m_0-{\hat \alpha}_+^*{\hat \alpha}_+
-{\hat \alpha}_-^*{\hat \alpha}_-}\ {\hat \alpha}_+ \ , \qquad
\label{4-11c}
\eeqn
\esub
On the other hand, the case $m_0=0$ gives us 
${\hat \alpha}_+^*{\hat \alpha}_+={\hat \alpha}_-^*{\hat \alpha}_-=0$. 
Therefore, the form (\ref{3-7}) reduces to the following form: 
\bsub\label{4-12}
\beqn
& &{\ovl I}_+
=-{\hat \beta}_+^*
\sqrt{m_1-{\hat \alpha}^*{\hat \alpha}-{\hat \beta}_+^*{\hat \beta}_+} \ , 
\qquad
{\ovl I}_-
=-\sqrt{m_1-{\hat \alpha}^*{\hat \alpha}-{\hat \beta}_+^*{\hat \beta}_+}\ 
{\hat \beta}_+ \ , 
\nonumber\\
& &{\ovl I}_0
={\hat \beta}_+^*{\hat \beta}_+
-(1/2)(m_1-{\hat \alpha}^*{\hat \alpha}) \ , 
\label{4-12a}\\
& &{\ovl M}_0
=(3/2){\hat \alpha}^*{\hat \alpha}-m_1/2 \ , 
\label{4-12b}\\
& &{\ovl D}_-^*
={\hat \alpha}^*{\hat \beta}_+ \ , 
\qquad
{\ovl D}_+^*
={\hat \alpha}^*
\sqrt{m_1-{\hat \alpha}^*{\hat \alpha}-{\hat \beta}_+^*{\hat \beta}_+} \ , 
\nonumber\\
& &{\ovl D}_-
={\hat \beta}_+^*{\hat \alpha} \ , 
\qquad
{\ovl D}_+
=\sqrt{m_1-{\hat \alpha}^*{\hat \alpha}-{\hat \beta}_+^*{\hat \beta}_+} \ 
{\hat \alpha}\ . 
\label{4-12c}
\eeqn
\esub
The above is identical with well-known Holstein-Primakoff representation 
in the case of the symmetric representation.

\section{The simplest approximation}

In order to illustrate that the Holstein-Primakoff representation presented in 
this paper is reasonable, we treat the simplest approximation which 
corresponds to the RPA method. 
Usually, the Hamiltonian of the $su(3)$-Lipkin model consists of two parts: 
${\hat H}={\hat H}_0+{\hat H}_1$. 
First part ${\hat H}_0$ is composed of the linear combination of 
${\hat I}_0$ and ${\hat M}_0$ and the second is expressed in terms of the 
bilinear form with respect to ${\hat I}_{\pm}$, ${\hat D}_{\pm}^*$ and 
${\hat D}_\pm$. 
In this paper, mainly we focus on ${\hat H}_0$, which can be expressed as 
\beqn\label{5-1}
{\hat H}_0&=&
(1/3)(\epsilon_0+\epsilon_1+\epsilon_2){\cal N} \nonumber\\
& &+(1/3)\left[(\epsilon_2-\epsilon_0)+(\epsilon_1-\epsilon_0)\right]{\hat M}_0
+(\epsilon_2-\epsilon_1){\hat I}_0 \ . 
\eeqn
Here, ${\cal N}$ denotes a $c$-number which corresponds to fermion number 
in the original fermion space and $\epsilon_0$, $\epsilon_1$ and 
$\epsilon_2$ represent the single-particle energies of the levels 0, 1 and 2, 
respectively. 
On the basis of the correspondence between the fermion and the Schwinger 
boson space discussed in Ref.\citen{4}, the form (\ref{5-1}) is derived from 
the fermion Hamiltonian ${\wtilde H}_0$: 
\begin{equation}\label{5-2}
{\wtilde H}_0=\epsilon_0{\wtilde N}_0+\epsilon_1{\wtilde N}_1+
\epsilon_2{\wtilde N}_2 \ .
\end{equation}
Here, ${\wtilde N}_0$, ${\wtilde N}_1$ and ${\wtilde N}_2$ denote the 
fermion number operators of the levels 0, 1 and 2, respectively. 
With the use of the relation (\ref{3-7}), ${\hat H}_0$ can be rewritten in 
the Holstein-Primakoff representation as follows: 
\beqn
& &{\ovl H}_0=U_0+{\ovl K}_0 \ , 
\label{5-3}\\
& &U_0=
(1/3)(\epsilon_0+\epsilon_1+\epsilon_2){\cal N} \nonumber\\
& &\qquad
-(1/3)\left[(\epsilon_2-\epsilon_0)+(\epsilon_1-\epsilon_0)\right]m_0
-(1/3)\left[(\epsilon_2-\epsilon_1)+(\epsilon_2-\epsilon_0)\right]m_1 \ , 
\label{5-4}\\
& &{\ovl K}_0=(\epsilon_2-\epsilon_0)({\hat \alpha}_+^*{\hat \alpha}_+
+{\hat \alpha}^*{\hat \alpha}) 
+(\epsilon_1-\epsilon_0){\hat \alpha}_-^*{\hat \alpha}_-
+(\epsilon_2-\epsilon_1){\hat \beta}_+^*{\hat \beta}_+ \ . 
\label{5-5}
\eeqn
We can see in the Hamiltonian ${\ovl K}_0$ that ``formally" there exist 
four types of the excitations.

Next, let us turn our eyes to the second part ${\hat H}_1$. 
In this paper, we do not contact this part concretely. 
The part ${\hat H}_1$ is expressed in terms of the bilinear form 
with respect to the operators ${\hat I}_\pm$, ${\hat D}_{\pm}^*$ and 
${\hat D}_\pm$. 
The equilibrium of our present system is characterized in terms of 
two parameters $m_0$ and $m_1$, and the fluctuations around 
the equilibrium are treated by the boson operators. 
As the simplest approximation, we pick up only the linear terms in the 
expressions (\ref{3-7}) and (\ref{3-8}) for the boson operators: 
\bsub\label{5-6}
\beqn
& &{\ovl I}_+=-\sqrt{m_1}\ {\hat \beta}_+^* \ , \qquad
{\ovl I}_-=-\sqrt{m_1}\ {\hat \beta}_+ \ , \nonumber\\
& &{\ovl D}_-^*=\sqrt{m_0}\ {\hat \alpha}_-^* \ , \qquad
{\ovl D}_-=\sqrt{m_0}\ {\hat \alpha}_- \ , 
\label{5-6a}\\
& &{\ovl D}_+^*=\sqrt{m_0}\ {\hat \alpha}_+^*+\sqrt{m_1}\ {\hat \alpha}^* \ , 
\qquad
{\ovl D}_+=\sqrt{m_0}\ {\hat \alpha}_++\sqrt{m_1}\ {\hat \alpha} \ , 
\label{5-6b}
\eeqn
\esub
\vspace{-1cm}
\begin{equation}\label{5-7}
{\check T}_+=\sqrt{m_0}\ {\hat \alpha}^*-\sqrt{m_1}\ {\hat \alpha}_+^* \ , 
\qquad
{\check T}_-=\sqrt{m_0}\ {\hat \alpha}-\sqrt{m_1}\ {\hat \alpha}_+ \ . 
\end{equation}
Since ${\ovl K}_0$ is of the bilinear with respect to the fluctuations, 
in our approximation, ${\ovl H}_1$ is also of the bilinear with respect to 
the fluctuations. 
This is in a form similar to the RPA method.

First, we consider the case $m_1=0$. 
In this case, we have 
\bsub\label{5-8}
\beqn
& &{\ovl I}_+=0 \ , \qquad
{\ovl I}_-=0 \ , 
\label{5-8a}\\
& &{\ovl D}_-^*=\sqrt{m_0}\ {\hat \alpha}_-^* \ , \quad
{\ovl D}_-=\sqrt{m_0}\ {\hat \alpha}_- \ , \quad
{\ovl D}_+^*=\sqrt{m_0}\ {\hat \alpha}_+^* \ , 
\quad
{\ovl D}_+=\sqrt{m_0}\ {\hat \alpha}_+ \ , \qquad
\label{5-8b}
\eeqn
\esub
\vspace{-1cm}
\begin{equation}\label{5-9}
{\check T}_+=\sqrt{m_0}\ {\hat \alpha}^* \ , 
\qquad
{\check T}_-=\sqrt{m_0}\ {\hat \alpha} \ . 
\hspace{6.5cm}
\end{equation}
The above relations tell us that the second part ${\ovl H}_1$ is expressed 
in terms of the bilinear form with respect to $({\hat \alpha}_{\pm}^*, 
{\hat \alpha}_\pm)$. 
Further, since ${\check T}_+{\check T}_-=m_0{\hat \alpha}^*{\hat \alpha}$ 
and 
$({\check T}_-\ket{ph}=0\ , \ \bra{ph}{\check T}_+=0)$, we can treat the 
system in the space with the number ${\hat \alpha}^*{\hat \alpha}=0$. 
Therefore, ${\ovl K}_0$ given in the relation (\ref{5-5}) can be expressed as 
\beqn\label{5-10}
{\ovl K}_0&=&
(\epsilon_2-\epsilon_0){\hat \alpha}_+^*{\hat \alpha}_+
+(\epsilon_1-\epsilon_0){\hat \alpha}_-^*{\hat \alpha}_- 
+(\epsilon_2-\epsilon_1){\hat \beta}_+^*{\hat \beta}_+ \ . 
\eeqn
For the expression (\ref{5-10}), we can give the interpretation that the 
first and the second terms represent the excitation from the level 0 to 2 
and 0 to 1, respectively, which appear in the conventional RPA. 
The third also appears in the conventional treatment and it troubles us. 
However, in our present case, the operator $(m_1-{\hat \alpha}^*{\hat \alpha}
-{\hat \beta}_+^*{\hat \beta}_+)$ appearing in the square root operator 
in the relation (\ref{3-7}) should be positive definite, and then, in the 
case $m_1=0$, not only ${\hat \alpha}^*{\hat \alpha}$ but also 
${\hat \beta}_+^*{\hat \beta}$ should vanish. 
Thus, we have 
\beqn\label{5-11}
{\ovl K}_0&=&
(\epsilon_2-\epsilon_0){\hat \alpha}_+^*{\hat \alpha}_+
+(\epsilon_1-\epsilon_0){\hat \alpha}_-^*{\hat \alpha}_- \ . 
\eeqn
In the case $m_0=0$, the treatment is in a form similar to the case $m_1=0$. 
We list up the relations in the following form: 
\bsub\label{5-12}
\beqn
& &{\ovl I}_+=-\sqrt{m_1}\ {\hat \beta}_+^* \ , \qquad
{\ovl I}_-=-\sqrt{m_1}\ {\hat \beta}_+ \ , 
\label{5-12a}\\
& &{\ovl D}_-^*=0 \ , \qquad
{\ovl D}_-=0 \ , \qquad
{\ovl D}_+^*=\sqrt{m_1}\ {\hat \alpha}^* \ , 
\qquad
{\ovl D}_+=\sqrt{m_1}\ {\hat \alpha} \ , \qquad
\label{5-12b}
\eeqn
\esub
\vspace{-0.9cm}
\beqn
& &{\check T}_+=-\sqrt{m_1}\ {\hat \alpha}_+^* \ , 
\qquad
{\check T}_-=-\sqrt{m_1}\ {\hat \alpha}_+ \ , 
\hspace{4.5cm}
\label{5-13}\\
& &{\ovl K}_0=(\epsilon_2-\epsilon_0){\hat \alpha}^*{\hat \alpha}
+(\epsilon_2-\epsilon_1){\hat \beta}_+^*{\hat \beta}_+ \ . 
\label{5-14}
\eeqn

In the case $(m_0\neq 0,\ m_1\neq 0)$, it may be convenient to introduce 
the following operators: 
\bsub\label{5-15}
\beqn
& &{\hat \delta}_+^*=\left(\sqrt{m_0+m_1}\right)^{-1}
\left(\sqrt{m_0}\ {\hat \alpha}_+^*+\sqrt{m_1}\ {\hat \alpha}^*\right) \ , 
\nonumber\\
& &{\hat \delta}_+=\left(\sqrt{m_0+m_1}\right)^{-1}
\left(\sqrt{m_0}\ {\hat \alpha}_++\sqrt{m_1}\ {\hat \alpha}\right) \ , 
\label{5-15a}\\
& &{\hat \delta}^*=\left(\sqrt{m_0+m_1}\right)^{-1}
\left(\sqrt{m_0}\ {\hat \alpha}^*-\sqrt{m_1}\ {\hat \alpha}_+^*\right) \ , 
\nonumber\\
& &{\hat \delta}=\left(\sqrt{m_0+m_1}\right)^{-1}
\left(\sqrt{m_0}\ {\hat \alpha}-\sqrt{m_1}\ {\hat \alpha}_+\right) \ . 
\label{5-15b}
\eeqn
\esub
Then, we have 
\begin{equation}\label{5-16}
{\hat \alpha}_+^*{\hat \alpha}_+ + {\hat \alpha}^*{\hat \alpha}
={\hat \delta}_+^*{\hat \delta}_+ + {\hat \delta}^*{\hat \delta} \ . 
\end{equation}
Further, in this case, the relations (\ref{5-6b}) and (\ref{5-7}) 
can be rewritten as 
\beqn
& &{\ovl D}_+^*=\sqrt{m_0+m_1}\ {\hat \delta}_+^* \ , \qquad
{\ovl D}_+=\sqrt{m_0+m_1}\ {\hat \delta}_+ \ , 
\label{5-17}\\
& &{\check T}_+=\sqrt{m_0+m_1}\ {\hat \delta}^* \ , \qquad
{\check T}_-=\sqrt{m_0+m_1}\ {\hat \delta} \ . 
\label{5-18}
\eeqn
Therefore, we describe the system in terms of 
$({\hat \alpha}_-^*,{\hat \alpha}_-)$, $({\hat \beta}_+^*,{\hat \beta}_+)$ 
and 
$({\hat \delta}_+^*,{\hat \delta}_+)$ and ${\ovl K}_0$ is expressed in the 
form 
\begin{equation}\label{5-19}
{\ovl K}_0=(\epsilon_2-\epsilon_0){\hat \delta}_+^*{\hat \delta}_+
+(\epsilon_1-\epsilon_0){\hat \alpha}_-^*{\hat \alpha}_- 
+(\epsilon_2-\epsilon_1){\hat \beta}_+^*{\hat \beta}_+\ . 
\end{equation}
We can see that the case $m_1=0$ and the case $m_0=0$ are described in 
terms of two kinds of the excitations, but, the case $(m_0\neq 0, m_1\neq 0)$ 
in terms of three kinds of the excitations.

Let us investigate the difference of the mechanism of these excitations. 
For this investigation, we must remember Ref.\citen{4}; 
the discussion on the correspondence between the original fermion and the 
Schwinger boson representation for the Lipkin model. 
The intrinsic state $\kket{m}$ in the original fermion space is 
specified by the occupation numbers for the level 0, 1 and 2, that is, 
$n_0=\Omega-n$, $n_1$ and $n_2$, respectively. 
Here, $\Omega$ denotes the degeneracy of each level. 
In Ref.\citen{4}, we showed the relation 
\begin{equation}\label{5-20}
n_1=n_0-m_0 \ , \qquad n_2=n_0-m_0-m_1 \ . 
\end{equation}
Then, if $m_1=0$, $n_1=n_2<n_0$ and $m_0$ fermions can excite from 
the level 0 to 1 or 2. 
If $m_0=0$, $n_2<n_1=n_0$ 
and $m_1$ fermions can excite from the level 0 or 1 to 
2. 
Therefore, these two cases can be described in terms of two kinds of bosons. 
In the case $(m_0\neq 0, m_1\neq 0)$, 
$m_0$ fermions, $m_1$ fermions and $(m_0+m_1)$ fermions can excite from 
the level 0 to 1, the level 1 to 2 and the level 0 to 2, respectively. 
Therefore, three kinds of boson operators are necessary. 
The above situation is interpreted in the original fermion space, 
in which the commutation relations are approximated as follows: 
\bsub\label{5-21}
\beqn
& &[\ {\wtilde I}_+\ , \ {\wtilde I}_-\ ]=2{\wtilde I}_0
\approx 2\bbra{m}{\wtilde I}_0\kket{m}=-m_1 \ , 
\label{5-21a}\\
& &[\ {\wtilde D}_-^*\ , \ {\wtilde D}_-\ ]={\wtilde M}_0-{\wtilde I}_0
\approx \bbra{m}{\wtilde M}_0-{\wtilde I}_0\kket{m}=-m_0 \ , 
\label{5-21b}\\
& &[\ {\wtilde D}_+^*\ , \ {\wtilde D}_+\ ]={\wtilde M}_0+{\wtilde I}_0
\approx \bbra{m}{\wtilde M}_0+{\wtilde I}_0\kket{m}=-(m_0+m_1) \ . 
\label{5-21c}
\eeqn
\esub
Here, ${\wtilde I}_{\pm}$, ${\wtilde D}_\pm^*$ and ${\wtilde D}_\pm$ 
denote the operators in the original fermion space which correspond to 
${\hat I}_\pm$, ${\hat D}_\pm^*$ and ${\hat D}_\pm$, respectively. 
In the present approximation, the other commutation relations vanish. 
The relations (\ref{5-21a}) $\sim$ (\ref{5-21c}) correspond to the 
relations (\ref{5-6a}) and (\ref{5-17}). 

The above argument may support that our present formalism is reasonable and 
it may serve in the study of the higher order corrections.

%



\end{document}